\documentclass[12pt]{article}
\def\sumk{ \sum_{k=1}^L  }
\def\half{{\textstyle {1\over 2}}}
\def\fourth{{\textstyle {1\over 4}}}
\def\eighth{{\textstyle {1\over 8}}}
\def\sixteenth{{\textstyle {1\over 16}}}
\def\e{{\rm e}}
\def\hO{{\hat O}}
\def\hQ{{\hat Q}}
\def\Tr{{\rm Tr}}
\def\SU{{\rm SU}}
\newcommand{\cA}{{\cal A}}
\newcommand{\cN}{{\cal N}}

\newcommand{\cP}{{\cal P}}
\newcommand{\cZ}{{\cal Z}}

\def\b{\beta}
\def\d{\delta} 
\def\la{\lambda} 

\def\be{\begin{equation}}
\def\ee{\end{equation}}
\def\bea{\begin{eqnarray}}
\def\eea{\end{eqnarray}}

\def\mininbar{\vrule height.75ex width.3pt depth0pt}
\def\cc{\relax\,\hbox{$\mininbar\kern-.2em{\hbox{\rm\tiny C}}$}}

\newcommand{\R}{\mathrm{I}\kern -2.5pt \mathrm{R}}
\newcommand{\Z}{\mathsf{Z}\kern -5pt \mathsf{Z}}
\newcommand{\1}{{1\kern -3pt \mathrm{l}}}

\begin{document}
\bibliographystyle{bst}

\begin{flushright} 
{\tt hep-th/0504222}\\ 
BRX-TH-562\\ 
BOW-PH-134\\
\end{flushright}
\vspace{30mm}

\vspace*{.3in}

\begin{center}
{\Large\bf More Pendants for Polya:}

{\Large\bf Two loops in the SU(2) sector}

\vskip 5mm
Marta~G\'omez-Reino\footnote{Research 
supported by the DOE under grant DE--FG02--92ER40706.}$^{,a}$,
Stephen G. Naculich\footnote{Research
supported in part by the NSF under grant PHY-0140281.}$^{,b}$,
and Howard J. Schnitzer\footnote{Research
supported in part by the DOE under grant DE--FG02--92ER40706.\\
{\tt \phantom{aaa} 
marta@brandeis.edu, 
schnitzer@brandeis.edu, 
naculich@bowdoin.edu}\\
\phantom{a;} ${}^*$ Permanent address.
}$^{,a*,c}$

\end{center}

\vskip 1mm

\begin{center}
$^{a}${\em Martin Fisher School of Physics\\
Brandeis University, Waltham, MA 02454}

\vspace{.2in}

$^{b}${\em Department of Physics\\
Bowdoin College, Brunswick, ME 04011}

\vspace{.2in}

$^{c}${\em Department of Physics\\
Harvard University, Cambridge, MA 02138}
\end{center}

\vskip 2mm

\begin{abstract} 
We extend the methods of Spradlin and Volovich
to compute the partition function for 
a conformally-invariant gauge theory on $\R \times S^3$
in which the dilatation operator is represented by a 
spin-chain Hamiltonian acting on pairs of states, 
not necessarily nearest neighbors.
A specific application of this is the two-loop dilatation operator 
of the planar SU(2) subsector of the 
$\cN=4$ SU($N$) super Yang-Mills theory in the large-$N$ limit. 
We compute the partition function and Hagedorn temperature 
for this sector to second order in the gauge coupling.
The Hagedorn temperature is to be interpreted as 
giving the exponentially-rising
portion of the density of states of the SU(2) sector,
which may be a signal of stringy behavior in the dual theory.

\end{abstract}

\vfil\break


\section{Introduction}
\setcounter{equation}{0}

The study of $\cN = 4$ supersymmetric Yang-Mills theory
in four dimensions has attracted a great deal of attention,
particularly motivated by the AdS/CFT correspondence, 
where the effort is to make a detailed comparison of 
the gauge theory and the dual string theory.
Several strategies are undergoing development,
one of which is inspired by the suggestion made in 
refs.~\cite{Berenstein:2002jq,Gubser:2002tv,
Frolov:2002av,Frolov:2003qc,Frolov:2003xy} 
to consider a subsector of semiclassical string states with large 
quantum numbers combined with the relation found in 
refs.~\cite{Minahan:2002ve,Beisert:2003tq,Beisert:2003yb} between the 
${\cN=4}$ supersymmetric Yang-Mills dilatation operator with the 
Hamiltonian of an integrable quantum spin chain. These facts allow 
for a precise comparison 
between the large-spin dependence of the string energies and the anomalous 
dimensions of gauge theory operators with large conformal dimensions. 
The spin chain Hamiltonian 
can also be mapped to a non-linear sigma model 
\cite{Kruczenski:2003gt,Kruczenski:2004kw,Hernandez:2004uw,Hernandez:2004kr}. 
(For reviews and further 
references see refs.~\cite{Tseytlin:2003ii,Beisert:2004ry,Tseytlin:2004xa}.)

Another approach to Yang-Mills theories in four dimensions 
(more relevant to the subject of this paper) is 
to consider a weakly-coupled 3+1 dimensional large-$N$ SU$(N)$ 
Yang-Mills theory compactified on a small $S^3$ and 
to study its large-$N$ thermodynamics.
For conformally-invariant theories, 
such as $\cN=4$ SYM theories,
the coupling constant does not run and may be taken
to be small independently of the volume of the space.
For asymptotically-free theories, 
such as pure SU$(N)$ YM theory,
the volume of the space provides an infrared cutoff 
on the running of the coupling constant. 
Thus the effective coupling may be made arbitrarily weak
at all energy scales for small volumes. This theory 
possesses very interesting thermodynamics \cite{Witten:1998zw} 
including a Hagedorn transition 
at zero 't Hooft coupling \cite{Sundborg:1999ue} 
(see also 
refs.~\cite{Aharony:2003sx,Aharony:2005bq,Liu:2004vy,Alvarez-Gaume:2005fv})
and a first-order deconfinement transition 
at weak coupling \cite{Aharony:2003sx,Aharony:2005bq}. 
(Similar behavior has also been found for other systems
\cite{Aharony:2004ig,Hadizadeh:2004bf}.)
This deconfinement transition, at least to lowest order,
is smoothed to third-order
in the presence of $N_f$ fields in the fundamental 
representation with $N_f/N$ 
finite \cite{Schnitzer:2004qt,Dumitru:2004gd}.

The partition function of the $\cN=4$ SU$(N)$ super Yang-Mills theory on
$\R \times S^3$ 
can be computed exactly at zero coupling 
in the $N\to\infty$ limit
by counting gauge-invariant operators using Polya theory 
\cite{Sundborg:1999ue,Aharony:2003sx,Sundborg:2000wp,Bianchi:2003wx}. 
One can also compute the perturbative corrections to the partition 
function at weak coupling.
It was shown in ref.~\cite{Spradlin:2004pp} how to 
compute the one-loop correction to the partition function below the Hagedorn 
transition (and from it the one-loop correction to the Hagedorn temperature)
by representing the 
one-loop dilatation operator as a periodic spin chain acting on 
nearest neighbors. 

In this paper, we continue the study of 
perturbative corrections to the partition function at non-zero Yang-Mills 
coupling. In order to do so, we extend the methods 
of Spradlin and Volovich \cite{Spradlin:2004pp} 
to compute the partition function for
a conformally-invariant gauge theory on $\R \times S^3$
in which the dilatation operator is represented by a 
spin-chain Hamiltonian acting on pairs of states, not
necessarily nearest neighbors.
A specific application of this is the two-loop dilatation
operator of the planar SU(2) subsector of the 
$\cN=4$ SU$(N)$ SYM theory in the large-$N$ limit. 
We compute the partition function and Hagedorn temperature 
for this sector to second order in the gauge coupling.
As the SU(2) sector is not closed at finite temperature, 
the Hagedorn temperature is to be interpreted as 
giving the exponentially-rising
portion of the density of states of the SU(2) sector,
and not the thermodynamics of the theory in a thermal bath.

The outline of the paper is as follows: in Section 2 we review  
the results of ref.~\cite{Spradlin:2004pp} that are relevant to our calculation.
In Section 3 we describe how to obtain the Hagedorn temperature
from the partition function computed in perturbation theory.
Section 4 describes the computation of the matrix elements
for a periodic spin-chain Hamiltonian with pairwise interactions
that do not necessarily act between nearest neighbors.
In Section 5 and Section 6, we specialize these results to the
SU(2) subsector in the planar limit. Finally in Section 7 we present 
some concluding remarks.

\section{Partition function of ${\cN}$=4 SYM on $\R \times S^3$}
\setcounter{equation}{0}

In this section, we review some of the results obtained in 
ref.~\cite{Spradlin:2004pp} for weakly coupled $\cN=4$ 
SYM theory on $\R \times S^3$ relevant to this paper.

Thermodynamics begins with consideration of the partition function 
\be
{\cal Z} ( \b ) = \Tr [ \e^{-\b H}] \,,
\ee
where $\b = 1/T$,
and $H$ is the Hamiltonian
of the $\cN=4$ $\SU(N)$ super Yang-Mills theory on $\R \times S^3$,
with the radius of $S^3$ taken to be unity.
(The dependence of physical quantities on radius can be restored
using dimensional analysis.)
Using the correspondence between
states of the YM theory on $\R \times S^3$
and gauge-invariant operators of the theory on $\R^4$
(where the Hamiltonian $H$ of the former
corresponds to the dilatation operator $D$ of the latter),
we rewrite the partition function as
\be
{\cal Z} (x) = \sum_{O} x^{D(O)} \,,
\ee
where $x = \e^{-\beta}$,
and the sum is over all gauge-invariant operators $O$ of
$\cN=4$ $\SU(N)$ Yang-Mills theory on $\R^4$.
The dilatation operator may be expanded perturbatively
in the 't Hooft parameter $\la = g^2_{YM} \: N$,
\be
\label{eq:dil}
D =  \sum^\infty_{n=0} \left[ \frac{\la}{ (4\pi) ^2 }\right]^n \:
D_{2n} \, .
 \ee
The zero-coupling partition function 
\be
{\cZ}^{(0)} (x)  = \sum_{O} x^{D_0 (O)}
\ee
may be computed by counting all the gauge-invariant 
operators of the theory weighted by $x^{D_0}$,
where $D_0$ 
just yields the engineering dimension of the operator $O$.  

The most general gauge-invariant operator is a linear combination
of $k$-trace operators, 
expressed as a product of $k$ single-trace operators,  
so a complete basis for gauge-invariant operators 
may be specified in terms of a complete basis of single-trace operators.  
Since single-trace operators have well-defined engineering dimension,
one may compute the zero-coupling partition function of 
this subset of operators:
\be
Z^{(0)} (x) = \Tr[ x^{D_0} ]\,,
\ee
where we will use $\Tr$ to denote the sum over single-trace operators.
The partition function of operators with an arbitrary number
of traces may then be written
\be
\label{eq:multi}
 {\cZ}^{(0)}  (x) 
= \exp \left[ \sum^\infty_{n=1} \frac{1}{n}\: Z^{(0)}(x^n)\right]
\ee
by treating the single-trace operators making up the general
multi-trace operators as indistinguishable bosons.

In turn, a general single-trace operator is 
an arbitrary product of the ``letters'' 
comprising the ``alphabet'' $\cal A$ 
of the ${\cN}$=4 SYM theory \cite{Polyakov:2001af}.
The counting of single-trace operators is equivalent
to the counting of the number of inequivalent ``necklaces'' $N$
constructed from beads of the alphabet $\cal A$ \cite{Sundborg:1999ue}.
The analogy of the necklace is appropriate since a necklace is
invariant under translation of the beads around the string,
just as a single-trace operator is invariant under cyclic permutation
of the fields composing it.
The zero-coupling single-trace partition function 
is therefore given by the necklace partition function
\be
\label{eq:necklace}
Z^{(0)} (x) = \sum_N \: x^{d(N)} \,, 
\ee
where each necklace is weighted by $d(N)$,
which is the sum of the engineering dimensions $d(A)$
of the beads $A$ belonging to it.
Equation (\ref{eq:necklace}) may be computed using Polya theory,
yielding \cite{Sundborg:1999ue}
\be
\label{eq:polya}
Z^{(0)} (x) 
=  \, - z(x) \, - \, \sum_{n=1}^\infty 
\frac{\phi(n)}{n} \ln \left[ 1- z(x^n) \right] \,, 
\ee
where 
\be
\label{eq:z}
 z (x) = \sum_{A \in {\cal A}} x^{d(A)} 
\ee
is the elementary partition function of the individual fields $A \in \cal A$,
and $\phi (n)$ is the Euler totient function,
denoting the number of integers less than $n$ 
that are relatively prime to $n$.
The term $-z(x)$ serves to remove the contribution 
of necklaces with a single bead,
since $\Tr[A]=0$ in $\SU(N)$.
The single-trace result (\ref{eq:polya}) may be inserted into
eq.~(\ref{eq:multi}) to yield \cite{Aharony:2003sx}
\be
\label{eq:hag}
 {\cZ}^{(0)}  (x) 
= \exp \left[ - \sum^\infty_{n=1} \frac{1}{n}\: z(x^n)\right]
\prod_{n=1}^\infty \frac{1}{1 - z(x^n)}\,.
\ee
The expressions above are valid only 
in the infinite-$N$ limit, 
since for finite $N$ there are relations among the traces
that reduce the number of independent operators.

Following ref.~\cite{Spradlin:2004pp}, we recast the computation of the 
necklace partition function in eq.~(\ref{eq:necklace}) as
a sum over periodic spin-chains of varying length $L$
\be
Z^{(0)} (x) 
= \sum^\infty_{L=1} \Tr_L [\cP x^d]\,,
\ee
where the spin vector at each site takes values $A \in \cA$ 
and is weighted by $x^{d(A)}$,
and $\Tr_L$ is the sum over spin chains of length $L$.
Here
\be
\cP = \cP^2 = \frac{1}{L} \; \sumk \; T^k
\ee
projects onto cyclically-invariant spin configurations,
where $T$ is the translation operator on the spin chain,
with the spin vector at site $i+L$ identified with that at site $i$.
Using this approach, one may rederive the result (\ref{eq:polya}) for the
zero-coupling partition function \cite{Spradlin:2004pp}.

Using this setup, 
the authors of ref.~\cite{Spradlin:2004pp} also computed 
the one-loop correction (i.e., first-order in $\lambda$)
to the partition function 
\bea
Z(x) 
&=& 
\Tr [x^{   D_0 + \{\la/(4\pi)^2\} D_2   + \cdots}]
\nonumber\\
\label{pf} &=& 
Z^{(0)} (x) 
+ \frac{\la}{(4\pi)^2}  \: ( \ln \, x) \: \Tr [x^{D_0} D_2] + \cdots
\eea
by counting necklaces with a ``pendant'' attached to 
two adjacent beads in a strand.
This is because the one-loop correction $D_2$ 
to the dilatation operator only acts on pairs of fields
adjacent to one another in the trace.
They computed this quantity by summing over spin chains
with insertions of an operator $\hO_1$
(related to $D_2$)
\be
\label{eq:Oone}
\Tr_{L} \left[ \cP x^d \hO_1 \right] ,
\qquad {\rm with} \qquad
\hO_1 = \sum_{i=1}^{L} \hO_{i,i+1}\,,
\ee
where $\hO_{i,i+1}$ acts non-trivially on the 
$i$th and $(i+1)$th spin variables, 
and trivially on the remainder of the spin variables.
Their result was \cite{Spradlin:2004pp}
\bea
\label{eq:SVresult}
\Tr_L[\cP x^d \hO_1] 
&=& \sumk \left[ z(x^{L/(k,L)}) \right]^{(k,L) - 2}
\langle O(x^{L/(k,L)}) \rangle
\cr
&+&
\sum_{k=1 \atop (k,L) = 1}^{L}
\left[
\langle PO(x^{L-k}, x^{k})\rangle - z(x^L)^{-1}
\langle O(x^L)\rangle\right]\,,
\eea
with
\bea
\label{eq:vevO}
\langle O(x) \rangle 
&=& \sum_{A_1,A_2} 
x^{d(A_1) + d(A_2)} 
\langle A_1 A_2 |\hO | A_1 A_2 \rangle  \; ,
\\
\label{eq:vevPO}
\langle PO (w,y) \rangle 
&=& \sum_{A_1,A_2} w^{d(A_1)} y^{d(A_2)} 
\langle A_1 A_2 | \hO | A_2 A_1 \rangle \; .
\eea
where $(k,L)$ denotes the greatest common divisor of $k$ and $L$.
Using eqs.~(\ref{eq:SVresult}),  (\ref{eq:vevO}), and (\ref{eq:vevPO}),
the authors of ref.~\cite{Spradlin:2004pp} 
explicitly evaluated the one-loop correction 
to the  partition function (\ref{pf})
for various sectors of the $\cN=4$ theory 
(see also ref.~\cite{Spradlin:2004sx}). 
In Section 4 we will show how one can generalize the approach above
in order to include operators which act pairwise on 
non-nearest neighbor spin variables.

\section{Hagedorn temperature}
\setcounter{equation}{0}

In this section, we describe how the perturbative computation of
the partition function (outlined in the previous section)
is used to compute the Hagedorn temperature of the theory.

Due to the exponentially-rising density of states
of the gauge theory, 
the partition function diverges at a finite
temperature, known as the Hagedorn temperature.
One may see from the $n=1$ term of the product in eq.~(\ref{eq:hag}) that 
the zero-coupling partition function goes as 
\be
\cZ^{(0)} (x) \sim  \frac{c_0}{x_0 - x}\,,
\ee
where $x_0$ is defined by  $z(x_0)=1$;  
the zero-coupling Hagedorn temperature is therefore 
$T_0 = -1/(\ln x_0)$.
The full partition function likewise diverges
\be
\label{eq:fullhagedorn}
\cZ (x) \sim  \frac{c}{x_H - x}\,,
\ee
but the location $x_H$ (and therefore the Hagedorn temperature)
is shifted from $x_0$ by higher-loop corrections.
Perturbatively expanding to second order in $\la$,
\be
x_H =  x_0 + \la x_1 + \la^2 x_2 + \cdots\;,
\qquad
c =  c_0 + \la c_1 + \la^2 c_2 + \cdots
\ee
we find
\be
\label{eq:hagexp}
\ln \left[ \frac{\cZ  (x)} {\cZ^{(0)}  (x) } \right]
 = \la   \left[  \left( \frac{x_1}{x - x_0} \right)
                 + \frac{c_1}{c_0} 
        \right]
 + \la^2 \left[ \frac{1}{2} \left( \frac{x_1}{x - x_0}\right)^2 
                         + \left( \frac{x_2}{x - x_0}  \right)
              + \left( \frac{c_2}{c_0} - \frac{c_1^2}{2 c_0^2} \right) 
        \right] + \cdots
\ee
Hence, the  corrections to the Hagedorn temperature may be
read directly from the coefficients of the simple poles of 
$\ln \left[ {\cZ  (x)}/ {\cZ^{(0)}  (x) } \right]$.

If we restrict ourselves to the planar sector of the theory,
we can neglect mixing between gauge-invariant operators with 
different trace structures \cite{Spradlin:2004pp}.
In this case, the basis for multi-trace operators can be
still be constructed from the basis of single-trace operators
and the full partition function written as 
\be
{\cZ  (x)}
= \exp \left[ \sum^\infty_{n=1} \frac{1}{n}\: Z(x^n)\right]\,,
\ee
just as in the zero-coupling case (\ref{eq:multi}).
{}From this, we have
\be
\label{eq:singletrace}
\ln \left[ \frac{\cZ  (x)} {\cZ^{(0)}  (x) } \right]
= \sum^\infty_{n=1} \frac{1}{n}\: \left[ Z(x^n) - Z^{(0)} (x^n) \right]\,,
\ee
so that (in the planar theory) the shift in 
Hagedorn temperature may be computed in terms of the
poles of the perturbative expansion of the 
single-trace partition function $Z(x)$.

We now expand the single-trace partition function 
to second order in the coupling constant:
\bea
\label{eq:twoloop}
Z(x) 
&=& 
\Tr [x^{   D_0 + \{\la/(4\pi)^2\} D_2 + \{ \la^2/(4\pi)^4\} D_4  + \cdots } ]
\nonumber\\
&=& 
Z^{(0)} (x) 
+ \frac{\la}{(4\pi)^2}  \: ( \ln \, x) \: \Tr [x^{D_0} D_2] \nonumber \\
&+& \frac{\la^2}{(4\pi)^4}\: (\ln \, x) \: \Tr [x^{D_0} D_4] 
+ \frac{\la^2}{ 2 (4\pi)^4}\: (\ln \, x)^2 \: \Tr [x^{D_0} D_2^2] 
+ \ldots
\eea
As mentioned in the previous section, 
the first-order correction (given by $ \Tr [x^{D_0} D_2] $)
was calculated in ref.~\cite{Spradlin:2004pp} 
and then used to compute the first-order shift in the Hagedorn temperature. 
In the next section, we generalize this analysis 
in order to be able to compute the  second-order corrections
$\Tr [x^{D_0} D_4] $ and $\Tr [x^{D_0} D_2^2] $,
and from this the second-order shift in Hagedorn temperature.

\section{More pendants for Polya}
\setcounter{equation}{0}

Higher-loop corrections to the dilatation operator act on pairs of fields
that are {\it not} adjacent to one another in the trace, 
so it is necessary to consider a generalization
of eq.~(\ref{eq:Oone}), namely
\be
\label{eq:Ocdef}
\Tr_L [\cP x^d \hO_c ], \qquad\qquad
\hO_c = \sum^L_{i=1} \hO_{i,i+c} \; ,
\ee
where the operator $\hO_{i,i+c}$
acts on pairs of {\it non-nearest} neighbor spins.
We assume that $c$ is not a multiple of $L$,
so that $i$ and $i+c$ are distinct.
Equation (\ref{eq:Ocdef})  
may be understood in terms of the counting of necklaces
with a pendant attached to two {\it non-adjacent} beads.
The calculation proceeds analogously to that in ref.~\cite{Spradlin:2004pp}
\bea
 \Tr_L [\cP\, x^d \, \hO_c ] 
& = &   L \, \Tr_L [ x^d \, \hO_{1,1+c} \, \cP ] 
 =     \sumk \Tr_L [x^d \hO_{1,1+c} T^k ] 
\nonumber \\
& = & \sumk \sum_{A{_1},\ldots, A_L} x^{\Sigma_i d(A_i)}
 \langle A_1 \ldots A_L |   \hO_{1,1+c} \: T^k | A_1 \ldots A_L \rangle
\nonumber \\
\label{eq:Oc}
& = & \sumk \sum_{A{_1},\ldots,A_L} x^{\Sigma_i d(A_i)} 
\langle A_1A_{1+c} | \hO | A_{k+1}A_{k+1+c}\rangle
 \prod^L_{j=2 \atop j \neq 1+c} \d_{A_j A_{j+k} } \, .
\eea
To evaluate this sum,
consider a product of delta functions that 
does not omit any terms,  $\prod^L_{j=1} \d_{A_jA_{j+k}}$.
This would weave the necklace into $m \equiv (k,L)$ strands.  
Now omit $\d_{A_1 A_{1+k}}$ and $\d_{A_{1+c} A_{1+c+k}}$ 
from the product of delta functions, 
as per eq.~(\ref{eq:Oc}).
This breaks either one strand or two strands, 
depending on whether $A_1$ and $A_{1+c}$ are on the same strand 
or not.  
They will be on the same strand if and only if 
$c$ is a multiple of $m$.
We therefore consider two separate cases,
depending on the value of $k$:\\

\noindent
Case (i): $c$ is not a multiple of $m = (L,k)$ \\

In this case, two separate strands are broken.
Since $A_1$ and $A_{1+c}$ are not on the same strand, 
the delta functions connect $A_{k+1}$ with $A_1$ and
$A_{k+1+c}$ with $A_{1+c}$ in two cycles consisting of $L/m$ terms.  
The remaining $m-2$ strands are also of length $L/m$.
Each strand contributes a factor of $x^{(L/m) \: d(A_i)}$.  
Hence
\bea
&&\sum_{A_1, A_{1+c}}
 x^{(L/m) [d(A_1) + d(A_{1+c})]} \, \langle A_1 A_{1+c} | \hO | A_1A_{1+c}
 \rangle
\left( \sum_{A_\ell} x^{(L/m) d (A_\ell)} \right)^{m-2}
\nonumber\\
&&\quad\quad=  \langle O (x^{L/m}) \rangle \left[z (x^{L/m})\right]^{m-2}\,,
\eea
where we have used eqs.~(\ref{eq:z}) and (\ref{eq:vevO}).\\

\noindent
Case (ii):  $c$ is a multiple of $m = (L,k)$ but not a multiple of $L$\\

In this case, $A_1$ and $A_{1+c}$ are on the same strand, 
so omitting $\d_{A_1 A_{k+1}}$ and $\d_{A_{1+c} A_{k+1+c}}$ 
breaks that strand into two pieces.  
The two segments have lengths $n$ and $(L/m) - n$, 
where $n$ is the smallest integer such that $ nk = c $  mod $L$.
Hence 
\bea
&&\sum_{A_{1},A_{1+c}}
 x^{nd(A_{1+c}) + [(L/m)-n]d(A_1)} 
\langle A_1 A_{1+c} | \hO | A_{1+c}A_1 \rangle
\left( \sum_{A_\ell} x^{(L/m) d (A_\ell)} \right)^{m-1}
\nonumber\\
&& \quad\quad
=  \langle PO (x^{(L/m)-n(k,L,c)} , x^{n(k,L,c)}  ) \rangle 
\left[z (x^{L/m})\right]^{m-1}\,,
\eea
where
\be
 n(k, \: L, \; c) 
  =  \min \{ n \geq 0 : \;\; nk = c \;{\rm mod} \;L \}
\ee
and we have used eqs.~(\ref{eq:z}) and (\ref{eq:vevPO}).\\

Combining the results of the two separate cases, we obtain
(for $c$ not a multiple of $L$)
\bea
\label{eq:ourresult}
\Tr_L[\cP x^d \hO_c] 
&=& \sumk \left[ z(x^{L/m}) \right]^{m-2}
\langle O(x^{L/m}) \rangle
\\
&+&
\sum_{k=1 \atop m | c}^{L}
\left[ z(x^{L/m}) \right]^{m-1}
\left[
\langle PO(x^{L/m - n(k,L,c)}, x^{n(k,L,c)})\rangle - z(x^{L/m} )^{-1}
\langle O(x^{L/m} )\rangle\right], 
\nonumber
\eea
which reduces to (\ref{eq:SVresult}) when $c=1$.
In eq.~(\ref{eq:SVresult}),
$n(k,L,1)$ has been replaced by $k$
because the set $\{ n (k,L,1)\}$ coincides with $\{ k: (k,L)=1 \}$.
This does not generally hold for $c \geq 2$.

{}From the last term in eq.~(\ref{eq:twoloop}),
one can see that we also need to evaluate
\be
\Tr_L[\cP x^d \hO^2_1]  \, .
\ee
Recalling that $\cP^2=\cP$ and $[\cP, \hO_1] = 0$, we find
\bea
\label{eq:osq}
&&\Tr_L[\cP x^d \hO^2_1]  
=  \Tr_L[x^d \hO_1 \cP \hO_1 \cP]   
=  L^2 \Tr_L[x^d \hO_{1,2} \cP \hO_{1,2} \cP]   \nonumber\\ [.12in]
&&=   L^2 \sum_{A{_1},\ldots, A_L} \sum_{B{_1},\ldots, B_L} 
x^{\Sigma_i d(A_i)} 
\langle A_1 \ldots A_L |   \hO_{1,2} \cP | B_1 \ldots B_L \rangle
\langle B_1 \ldots B_L |   \hO_{1,2} \cP | A_1 \ldots A_L \rangle
\nonumber\\
&&=  
\sum^{L}_{k=1} 
\sum^{L}_{\ell=1} 
\sum_{A{_1},\ldots, A_L} \sum_{B{_1},\ldots, B_L} 
x^{\Sigma_i d(A_i)} 
\langle A_1 A_{2} |   \hO | B_{1+k-\ell} B_{2+k-\ell} \rangle \times
\nonumber\\
&&
\qquad \qquad \qquad \qquad \qquad \qquad \qquad
 \langle B_1 B_{2} |   \hO | A_{1+\ell} A_{2+\ell} \rangle
 \prod^L_{i=3} \d_{A_i B_{i+k-\ell} }
 \prod^L_{j=3} \d_{B_j A_{j+\ell} } \,.
\eea

As before, to evaluate this sum,
we will initially ignore the restrictions
$i$,$j  \neq 1$,$2$  on the delta functions.
The product of delta functions 
$ \prod^L_{i=1} \d_{A_i B_{i+k-\ell} } \prod^L_{j=1} \d_{B_j A_{j+\ell} } $
then weaves a necklace consisting of 
two sets of beads $\{A_i\}$ and $\{B_i\}$, 
connected  $A \to B \to A \to B \cdots$.
This necklace contains $m=(k,L)$ separate strands,
each of length $2L/m$.
The omission of the delta functions 
$\d_{A_1 B_{1+k-\ell}}$,
$\d_{A_{2} B_{2+k-\ell}}$,
$\d_{B_1 A_{1+\ell}}$, and 
$\d_{B_{2} A_{2+\ell}}$ 
has the effect
of breaking one, two, three, or four strands (depending on the values
of $k$ and $\ell$),
with the rest remaining intact.

It will turn out that, for the purpose of evaluating
the shift in the Hagedorn temperature, we will only need
the $k=L$ (mod $L$) term in the sum (\ref{eq:osq}),
so we now focus on this case.
In this case, we begin with a necklace with $L$ separate strands,
each consisting of an $A$ bead and  $B$ bead.
When $L\ge 3$,
the omission of the delta functions 
$\d_{A_1 B_{1-\ell}}$,
$\d_{A_{2} B_{2-\ell}}$,
$\d_{B_1 A_{1+\ell}}$, and 
$\d_{B_{2} A_{2+\ell}}$ 
has the effect
of breaking 
two (when $\ell=L$),
three (when $\ell= 1$ or $L-1$), or 
four (the other $L-3$ cases) strands,
the rest remaining intact.
Hence, the sum (\ref{eq:osq}) reduces to 
\bea
\label{eq:kzero}
\Tr_{L \ge 3} [\cP x^d \hO^2_1]   \bigg|_{k=L}
&=& \sum_{A_1,  A_2, B_1, B_2} x^{d(A_1)+ d(A_{2})} 
\Big[
 \langle A_1 A_2 |   \hO | B_1 B_2 \rangle
 \langle B_1 B_2 |   \hO | A_1 A_2 \rangle
z(x)^{L-2} \qquad\qquad
\nonumber\\
&& 
\quad \qquad \qquad
 ~+~
x^{d(B_2)}  \, 
 \langle A_1 A_2 |   \hO | A_1 B_1 \rangle
 \langle B_1 B_2 |   \hO | A_2 B_2 \rangle
z(x)^{L-3}
\nonumber\\ [.12in]
&& 
\quad \qquad \qquad
 ~+~
x^{d(B_1)}   \,
 \langle A_1 A_2 |   \hO | B_2 A_2  \rangle
 \langle B_1 B_2 |   \hO | B_1 A_1  \rangle
z(x)^{L-3}
\nonumber\\ [.12in]
&+&
(L-3)
\quad x^{d(B_1)+d(B_2)}   \,
\, \langle A_1 A_2 |   \hO | A_1 A_2 \rangle
 \langle B_1 B_2 |   \hO | B_1 B_2 \rangle
z(x)^{L-4}
\Big] \, .
\eea
The case $L=2$ differs slightly, yielding
\bea
\label{eq:kzeroLtwo}
\Tr_2[\cP x^d \hO^2_1]   \bigg|_{k=L}
&=& \sum_{A{_1},  A_{2}, B{_1}, B_{2}} x^{d(A_1)+ d(A_{2})} 
\Big[
\langle A_1 A_{2} |   \hO | B_{1} B_{2} \rangle
 \langle B_1 B_{2} |   \hO | A_{1} A_{2} \rangle
\qquad\nonumber\\
&&\quad\quad \qquad \qquad \qquad   ~+~
\langle A_1 A_{2} |   \hO | B_{2} B_{1} \rangle
 \langle B_1 B_{2} |   \hO | A_{2} A_{1} \rangle
\Big] \, .
\eea
We will evaluate these expressions explicitly in the following section
in the case of the SU(2) spin chain.

\section{Specialization to the SU(2) spin chain}
\setcounter{equation}{0}

Consider a restriction of the alphabet $\cA$ to two of the
complex scalars of the Yang-Mills theory, $X$ and $Z$,
which have $d(X) = d(Z) = 1$.
This corresponds to an SU(2) spin-chain.
For this sector, $z(x)=2x$, 
so from eq.~(\ref{eq:hag}), one observes
that the zero-coupling Hagedorn temperature
occurs at $x_0=\half$ or $T_0  =  1 /  (\ln 2)$.

For this sector, the dilatation operator can be
expressed in terms of the operator
\be
\hQ_{i,j} = \left( {\1} - P \right)_{ij}\,,
\ee
where $P_{ij}$ simply transposes two spin vectors:
\be 
P_{ij}  | \cdots A_i \cdots A_j \cdots \rangle 
  =  | \cdots A_j \cdots A_i \cdots \rangle\,. 
\ee
In this case one easily computes
\be
\langle Q (x)\rangle = 2x^2, \qquad
\langle PQ (w,y)\rangle = -2wy \; .
\ee
Inserting these in eq.~(\ref{eq:ourresult}), one obtains
\be
\label{eq:sutwo}
\Tr_L[\cP x^d \hQ_c] 
= \frac{x^L}{2}
\left[
	 \sumk 		2^{m}
     - 3 \sum_{k=1 \atop m | c}^{L} 	2^{m}
\right] \, , \qquad 
\hQ_c = \sum_{i=1}^{L} \hQ_{i,i+c}\,,
\ee
valid for $c$ not a multiple of $L$.
When $c$ is a multiple of $L$, $i$ and $i+c$ are identified,
so $\hQ = 0$ and 
$\Tr_L[\cP x^d \hQ_c] $
vanishes identically.

Next we sum $\Tr_L[\cP x^d \hQ_c] $ over all $L$, 
omitting terms $L$ on the r.h.s. of eq.~(\ref{eq:sutwo})
corresponding to divisors of $c$: 
\be
\label{eq:special}
\sum_{L=1}^\infty \Tr_L[\cP x^d \hQ_c] 
= 
\sum_{L=1}^\infty  
 \frac{x^L}{2}
\left[ \sumk 		2^{m}
     - 3 \sum_{k=1 \atop m | c}^{L} 	2^{m} \right]
- \sum_{L=1 \atop L | c }^\infty  
 \frac{x^L}{2}
\left[ \sumk 		2^{m}
     - 3 \sum_{k=1 \atop m | c}^{L} 	2^{m} \right] \, .
\ee
The sum over 
$k=1, \ldots, L$ of any summand that
only depends on $m=(k,L)$ 
may be rewritten as a sum over the divisors $m$ of $L$, 
or equivalently as the sum over the divisors $n=L/m$ of $L$: 
\be
\sumk  f(L, m) =
\sum_{m|L} \phi(L/m) f(L, m) =
\sum_{n|L} \phi(n) f(L, L/n) .
\ee
This allows us to rewrite eq.~(\ref{eq:special}) as
\bea
\label{eq:rewrite}
\sum_{L=1}^\infty \Tr_L[\cP x^d \hQ_c] 
&=& 
\sum_{L=1}^\infty  
 \frac{x^L}{2}
\left[ \sum_{n|L} \phi(n) 2^{L/n}
     - 3 \sum_{m|L \atop m|c} \phi({L}/{m})  2^{m} \right]
\nonumber\\
&-& \sum_{L | c }
 \frac{x^L}{2}
\left[ \sum_{n|L} \phi(n) 	2^{L/n}
     - 3 \sum_{n|L \atop (L/n) | c} \phi(n) 2^{L/n} \right]\,.
\eea
In the first term, $n$ divides $L$, so we may rewrite that term 
as an unrestricted sum over $n$ and $m$, with $L=nm$,
\be
\sum_{L=1}^\infty  
 \frac{x^L}{2} \sum_{n|L} \phi(n) 2^{L/n}
= \sum_{n=1}^\infty \frac{1}{2} \phi(n) 
\sum_{m=1}^\infty  (2 x^n )^m
= \sum^\infty_{n=1} \phi (n) \frac{x^n}{(1-2x^n)}  \, .
\ee
In the second term of eq.~(\ref{eq:rewrite}), $m$ divides $L$, 
so the sum over $L$ may be replaced by a sum over $n$ with $L=nm$,
and the restriction $m|L$ dropped.
In the fourth term of eq.~(\ref{eq:rewrite}), 
we have three restrictions:
$L|c$, $n|L$, and $(L/n) | c$. 
However, $n|L$ implies $(L/n) | L$, 
and since $L|c$, the third restriction may be dropped. 
Putting everything together, we obtain
\be
\label{eq:Qc}
 \sum^\infty_{L=1} \Tr_L [\cP x^d \hQ_c]
  =  \sum^\infty_{n=1} \phi (n) \frac{x^n}{(1-2x^n)}
   - \frac{3}{2} \: \sum_{m|c} 2^m\sum^\infty_{n=1} \phi (n) x^{mn}
   + \sum_{L|c} x^L \sum_{n|L} \phi (n) 2^{L/n} \, .
\ee
When $c=1$, this reduces to\footnote{The 
$L=1$ term on the l.h.s. vanishes identically.}
\be
 \sum^\infty_{L=1} \Tr_L [\cP x^d \hQ_1]
={2} 
\left[ x  - 
\sum^\infty_{n=1} \phi (n) x^n \left( \frac{1-3x^n}{1-2x^n} \right)\right]\,,
\ee
in agreement with eq.~(4.4) of ref.~\cite{Spradlin:2004pp}.
Below we will also need the $c=2$ case\footnote{The
$L=1$ and 2 terms on the l.h.s. vanish identically.}
\be
 \sum^\infty_{L=1} \Tr_L [\cP x^d \hQ_2]
={2} 
\left[ x   + 3 x^2
- \sum^\infty_{n=1} \phi (n) x^n \left( \frac{1-6x^{2n}}{1-2x^n} \right)\right]\,.
\ee

Let us also explicitly evaluate, 
for the SU(2) sector, the expressions appearing 
in eqs.~(\ref{eq:kzero}) and (\ref{eq:kzeroLtwo}):
\bea
\Tr_{L\ge 3} [\cP x^d \hQ^2_1]   \bigg|_{k=L}
&=& \left[ 4x^2 (2x)^{L-2} 
+ 2 \cdot 2x^2 (2x)^{L-3} 
+ (L-3) \cdot 4x^2 (2x)^{L-4} \right] 
= \frac{L+3}{4} (2x)^L,
\nonumber\\
\Tr_2[\cP x^d \hQ^2_1]   \bigg|_{k=L}
&=& \left[ 4x^2 + 4 x^2 \right] = 8 x^2 \,.
\eea
Summing these up, we obtain\footnote{The 
$L=1$ term on the l.h.s. vanishes identically.}
\be
\label{eq:crude}
\sum_{L=1}^\infty
\Tr_L[\cP x^d \hQ^2_1]   \bigg|_{k=L}
=    \frac{ 4 x^2 (1-x)(2-3x)} { (1-2x)^2 }\,,
\ee
which contains simple and double poles at $x=\half$,
the zero-coupling Hagedorn temperature.

We will now justify why the $k\neq L$ mod $L$ terms in eq.~(\ref{eq:osq})
will not be needed.
In the general case, the necklace contains $m = (k,L)$ strands,
each of length $2L/m$.
Before cutting the strands, the partition function
contains a term $z(x^{L/m})^m = 2^m x^L$.
Cutting the strands alters this slightly, but the
large $L$ asymptotic behavior remains the same.
In the subsequent sum over $L$, 
the terms $2^m x^L$ yield poles at $x = (1/2)^{m/L}$,
but the sum is convergent at smaller values of $x$.
In particular, the sum is finite at $x=\half$, except
for $m=L$ which only occurs when $k=L$ mod $L$,
as we saw in eq.~(\ref{eq:crude}).
Since, as we saw in eq.~(\ref{eq:hagexp}),
only poles at $x=\half$ contribute
to the shift of the Hagedorn temperature, 
we may neglect the terms with $k \neq L$ mod $L$
for this purpose.

\section{Two-loop partition function for the SU(2) sector}

We now have the ingredients necessary
to compute the two-loop correction to the
single-trace partition function in the planar
SU(2) sector of the SYM theory.
In this sector, the dilatation operator (\ref{eq:dil})
has the one- and two-loop 
corrections \cite{Minahan:2002ve, Beisert:2003tq,Ryzhov:2004nz}
(in terms of the SU(2) spin-chain)
\be
D_2 =  2 \hQ_1 \, , \qquad
D_4 =  2 \hQ_{2}  - 8 \hQ_{1} \,,
\ee
hence the two-loop partition function (\ref{eq:twoloop}) 
may be rewritten in terms of spin-chain observables as
\bea
\label{eq:twoloopspin}
Z(x) 
&=& 
Z^{(0)} (x) 
+ \frac{2 \la}{(4\pi)^2}  \: ( \ln \, x) \: 
 \sum^\infty_{L=1} \Tr_L [\cP x^d \hQ_1] \nonumber\\
&+& \frac{ 2 \la^2}{(4\pi)^4}\: (\ln \, x) \:  
\sum^\infty_{L=1} \Tr_L [\cP x^d \hQ_2] 
 -  \frac{ 8 \la^2}{(4\pi)^4}\: (\ln \, x) \:  
\sum^\infty_{L=1} \Tr_L [\cP x^d \hQ_1] 
\nonumber\\
&+& \frac{2 \la^2}{ (4\pi)^4}\: (\ln \, x)^2 \: 
\sum^\infty_{L=1} \Tr_L [\cP x^d \hQ^2_1]
+ \ldots
\eea

For the purposes of the computing the shift in
the Hagedorn temperature, 
we only require the coefficients of the poles at $x=x_0=\half$.
{}From eq.~(\ref{eq:Qc}),  one finds that 
\be 
\label{eq:Qcpole}
\sum^\infty_{L=1} \Tr_L [\cP x^d \hQ_c]  \,  
= \, \frac { -\fourth } { (x - \half) } + \cdots
\ee
As we saw in the previous section,
the contributions of $ \Tr_L [\cP x^d \hQ_1^2]  $
to poles at $x=\half$ arise only from the $k=L$ terms in the sum 
\be
\label{eq:Qonesqpole} 
\sum^\infty_{L=1} \Tr_L [\cP x^d \hQ^2_1]
\sim 
\sum^\infty_{L=1} \Tr_L [\cP x^d \hQ^2_1] \bigg|_{k=L}
\sim  
   \frac { \sixteenth  } { (x - \half )^2}
-  \frac { \fourth } { (x -\half )} + \cdots
\ee
Using 
eqs.~(\ref{eq:twoloopspin}), (\ref{eq:Qcpole}), and (\ref{eq:Qonesqpole}),
we have 
\bea
Z(x) -  Z^{(0)} (x)  
&\sim & \frac{2 \la}{(4\pi)^2}  \: ( \ln \, x) \: 
\left[ \frac{ - \fourth}{x - \half}  \right]
-  \frac{ 6 \la^2}{(4\pi)^4}\: (\ln \, x) \:  
\left[ \frac{- \fourth}{x-\half}  \right] 
\\
&+& 
\frac{2 \la^2}{ (4\pi)^4}\: (\ln \, x)^2 \: 
\left[ 
 \frac { \sixteenth  } { (x - \half )^2}
- \frac { \fourth } { (x -\half )} 
 \right]
+ \ldots
\nonumber\\
&=& \frac{\la}{(4\pi)^2}  
\left[ \frac{ \half \ln 2 }{\left(x -  \half \right)}  + \cdots \right]
+\frac{\la^2}{(4\pi)^4}  
\left[ \frac{ \eighth (\ln 2)^2 }{(x- \half)^2}  
	     - \frac{ 2 \ln 2 + \half (\ln 2)^2 } {(x - \half )} 
 + \cdots \right] + \cdots
\nonumber
\eea

Next we use this in eq.~(\ref{eq:singletrace})
and compare with eq.~(\ref{eq:hagexp}).
The double pole in the $\la^2$ term is consistent with eq.~(\ref{eq:hagexp}),
and the residues yield
the one- and two-loop corrections to the Hagedorn temperature
for the planar SU(2) sector:
\bea
x_H &=& 
\frac{1}{2}
+ \frac{\la} {(4\pi)^2}  \frac{\ln 2}{2}
- \frac{\la^2 }{(4 \pi)^4}
\frac{ \ln 2}{2}  \left(4 + \ln 2\right)
+ \cdots \, ,
\nonumber\\
\frac{T_H}{ T_0} &=& 1  + \frac{\la} {(4\pi)^2}  
- \frac{\la^2 }{(4 \pi)^4} \frac{3}{2}  \left(2 + \ln 2\right) + \cdots
\eea
Thus, the second-order correction to the Hagedorn temperature,
which represents the position of the exponentially-rising 
density of states of the planar SU(2) sector, is in the
opposite direction to the first-order shift. 
It should be noted  that in this case the Hagedorn temperature does not give 
the thermodynamics of the theory in a thermal bath 
(due to the fact that the SU(2) sector is not closed 
at finite temperature), 
but should be interpreted as 
giving the exponentially-rising
portion of the density of states of the SU(2) sector.

\section{Concluding remarks}
\setcounter{equation}{0}

In this paper, we have extended the methods of ref.~\cite{Spradlin:2004pp} 
to compute the partition function of a conformally-invariant
gauge theory on $\R \times S^3$ 
for which the dilatation operator acts on 
pairs of fields not necessarily adjacent to one another
in gauge-invariant operators.
We restricted the calculation to the planar large-$N$ limit
to avoid mixing between single and multi-trace operators.
The partition function of single-trace operators was 
evaluated using a periodic spin-chain 
whose Hamiltonian contains non-nearest neighbor interactions.

We specialized this to the SU(2) sector
of the $\cN=4$ SYM theory, 
and computed the partition function to second order in the gauge coupling.
This calculation does not represent the partition function
in a thermal bath, since the SU(2) sector is not closed
at finite temperature, but rather is to be interpreted as 
giving the density of states belonging to the Hamiltonian of this sector.
We computed the first- and second-order shift in the Hagedorn
temperature, which corresponds to the exponential rise
of the density of states of the SU(2) sector.
It is often speculated that this exponential rise is
associated with stringy behavior,
and is believed to continue to hold at strong coupling.

All the calculations in this paper have been done
at weak coupling.
The behavior of the theory could change dramatically at
strong coupling.
For example, the residue $c(\lambda)$ in eq.~(\ref{eq:fullhagedorn})
could vanish for some finite value of $\lambda$,
indicating a softer than exponential growth of the density of
states as the phase transition is approached.
Alternatively, the value of $x_H(\lambda)$ could approach 1, 
implying that a Hagedorn transition
is not attained at any finite temperature.

Finally, although we focused in this paper on (a particular sector of)
the $\cN=4$ SYM theory, 
the methods developed could be used for any
theory (or sector of a theory) in which the 
dilatation operator (at some order in perturbation theory) 
acts on pairs of fields not necessarily adjacent to 
one another in the gauge-invariant operator. Also, due to the 
connection between $\cN=4$ supersymmetric Yang-Mills theories 
and integrable systems, it would be very 
interesting to understand the role of integrability in these
types of calculations.

\section*{Acknowledgments}

We wish to thank Martin Kruczenski,
Joe Minahan, Shiraz Minwalla, and Anton Ryzhov 
for helpful conversations.
HJS wishes to thank the string group 
at the Harvard University physics department 
for their warm hospitality during his sabbatical leave.

\providecommand{\href}[2]{#2}\begingroup\raggedright\endgroup

\end{document}